\documentclass[preprint,showpacs,preprintnumbers,amsmath,amssymb]{revtex4}
\usepackage{graphicx}

\begin{document}

\thispagestyle{empty}

\title{The Casimir force between a microfabricated elliptic cylinder
and a plate}

\author{R.~S.~Decca,${}^1$ E.~Fischbach,${}^2$
G.~L.~Klimchitskaya,${}^3$
D.~E.~Krause,${}^{4,2}$ D.~L\'{o}pez,${}^5$
and V.~M.~Mostepanenko${}^6$
}

\affiliation{
${}^1$Department of Physics, Indiana University-Purdue
University Indianapolis, Indianapolis, Indiana 46202, USA\\
${}^2$Department of Physics, Purdue University, West Lafayette, Indiana
47907, USA\\
${}^3$North-West Technical University, Millionnaya Street 5, St.Petersburg,
191065, Russia \\
${}^4$Physics Department, Wabash College, Crawfordsville, Indiana 47933,
USA\\
${}^5$Center for Nanoscale Materials, Argonne National Laboratory,
Argonne, Illinois 60439, USA \\
${}^6$Noncommercial Partnership
``Scientific Instruments'',  Tverskaya Street 11, Moscow,  103905, Russia
}

\begin{abstract}
We investigate the Casimir force between a microfabricated
elliptic cylinder (cylindrical lens) and a plate made of real
materials. After a brief discussion of the fabrication procedure,
which typically results in elliptic rather than circular
cylinders, the Lifshitz-type formulas for the Casimir force and
for its gradient are derived. In the specific case of equal
semiaxes, the resulting formulas coincide with those derived
previously for circular cylinders. The nanofabrication procedure
may also result in asymmetric cylindrical lenses obtained from
parts of two different cylinders, or rotated
through some angle
about the axis of the cylinder. In these cases the
Lifshitz-type formulas for the Casimir force between
a lens and a plate and for its gradient are also derived, and
the influence of lens asymmetry is determined.
Additionally, we obtain an expression for the shift of the
natural frequency of a micromachined oscillator with an
attached elliptic cylindrical lens interacting with a plate
via the Casimir force in a nonlinear regime.
\end{abstract}

\maketitle

\section{Introduction}

In recent years the Casimir effect \cite{1} is acknowledged to be
among the most rapidly developing fields of fundamental physics.
It has attracted considerable attention as a test for the structure
of the quantum
vacuum, and for hypothetical interactions predicted in many extensions
of the standard model, and also opened up new opportunities for
nanotechnology \cite{2}. Since 1997 approximately 30 experiments
on measuring the Casimir force have been performed (see reviews
\cite{3,4}), which not only confirmed the currently available
theoretical knowledge, but also led to unexpected results of major
importance. Specifically, it was recognized \cite{5} that the
unified theory of van der Waals and Casimir forces developed by
Lifshitz encounters problems in the description of free charge
carriers. As a result, two theoretical approaches were proposed
based on the Drude \cite{6,7,8} and plasma \cite{9,10,11}
models. Lifshitz theory, combined with the seemingly most natural
Drude model, was shown to be in contradiction with the Nernst
heat theorem \cite{2,12,13,14} and with the experimental
data \cite{15,16}. In contrast Lifshitz theory using the plasma model
for the dielectric permittivity was found to be thermodynamically
and experimentally consistent, despite the fact that it does not
take into account the relaxation properties of free charge
carriers.
Note that the experiment of Refs.~\cite{15,16} is an independent
measurement of the gradient of the Casimir force with no fitting
parameters, such as a distance offset, etc. In the first repetition
of this experiment \cite{16a} measurements were performed at
separations up to 1.15$\,\mu$m where zero values of the force were
achieved within the limits of experimental errors.
It was shown \cite{16b} that the introduction of an offset did
not improve the agreement between data and the Drude model and
made
the agreement with the plasma model worse. Because anomalous
electrostatic contributions were not observed, introducing
additional parameters in the theoretical description of the
experimental data was unwarranted.

This situation has been the subject of much
controversy (see, for instance, Refs.~\cite{17,18,19,20,21}).
Along with the experimental and theoretical investigations
mentioned above, great progress was achieved in the
calculation of the Casimir force between nonplanar surfaces
based on the scattering approach \cite{2,22,23,24,25,25a,26}.
Bearing in mind, however, that in the end the elements of
a scattering matrix are expressed in terms of dielectric
permittivity or some other quantity characterizing
material properties of the test bodies, successful application
of new methods calls for the resolution of the problem of
free charge carriers.

Presently great interest is expressed in new measurements
of the Casimir force which could shed light on this problem.
Thus, the experiment \cite{27} claims observation of the
thermal Casimir force, as predicted by the Drude model approach,
in the separation region from 0.7 to $7\,\mu$m.
It should be mentioned, however, that in Ref.~\cite{27}
what is measured is
not only the thermal Casimir force, but
up to an order of magnitude greater total force presumably
determined by large surface patches.
The theoretical expression for the total force contains two
fitting parameters determined from the best fit between
the experimental data and theory. Therefore, Ref.~\cite{27}
is not an independent measurement of the Casimir force
as is the experiment of Refs.~\cite{15,16}.
In addition, it was shown \cite{28} that the simplest
version of the proximity force approximation (PFA), used  in
Ref.~\cite{27} to calculate both the Casimir and electric
force between a spherical lens with $R=15.6\,$cm radius
of curvature and a plate, is inapplicable to large lenses
due to the presence of surface imperfections.
Another recent experiment employing large spherical
lenses \cite{29} does not support the existence of a large
thermal correction to the Casimir force predicted by the
Drude model approach.
Because of this, new experiments, especially exploiting
more sophisticated configurations than a sphere or
a spherical lens above a plate, may lead to more
reliable results than those obtained in Refs.~\cite{27,29}.

As a prospective alternative configuration for the
measurement of the Casimir force, a cylinder-plate
geometry has long been discussed in the literature
\cite{30,31,32}.
This geometry is intermediate between the configurations of
two parallel plates and a sphere above a plate.
It preserves some advantages of the latter while making
the problem of preserving the parallelism less difficult
than for two plates.
However, the configuration of
cylinders with centimeter-size radii of curvature
revealed anomalies in electrostatic calibrations \cite{32}
which might be caused by surface imperfections.
To avoid this problem, Ref.~\cite{33} proposed an
experiment measuring the thermal Casimir interaction
between a plate and a microfabricated cylindrical lens
attached to a micromachined oscillator.
Such metallic lenses, with smooth surfaces of about
$100\,\mu$m radii of curvature on top of a micromachined
oscillator, can be directly fabricated by using a
monolithic fabrication process. In Ref.~\cite{33} the
Lifshitz-type formulas for the thermal Casimir force
between a circular cylinder and a plate made of real
metals were derived using the PFA.
{}From a comparison with exact results available for
ideal metals it was shown that for reasonable experimental
parameters the error resulting from the use of the PFA
is much less than 1\%.
This conclusion was confirmed in Ref.~\cite{34a} for an
ideal metal cylinder above an ideal metal plate.
It was shown that in the region of experimental
temperatures the PFA correctly reproduces the dominant
contributions to both the Casimir force and thermal
correction to it. The validity of the PFA was also
confirmed \cite{34b} for the configuration of an atom near an ideal
metal cylinder.
Reference \cite{33} demonstrated the feasibility of the
proposed experiment, and investigated corrections to the
Casimir force and its gradient due nonparallelity of a
plate and a cylinder axis, and due to the finite length
of a cylinder.

In this paper we investigate the Casimir force between
a microfabricated elliptic cylinder and a plate using
the PFA approach. Our consideration is adapted to the
measurement scheme using a micromachined oscillator.
Motivation for use of elliptic cylinders derives from the
fact that fabrication procedures usually result in
cylinders with semiaxes in two perpendicular directions
varying by 20\%--30\%. Fabrication may result also
in asymmetric cylindrical lenses consisting of parts
of two different elliptic cylinders or rotated through
some angle about the cylinder axis. In all these
cases we derive the Lifshitz-type
formulas for the Casimir force
and for its gradient and perform computations
to account for the
role of asymmetry. The electric force between an elliptic
cylindrical lens and a plate is also calculated for the
purpose of electrostatic calibration of the Casimir setup.
Furthermore, we consider an elliptic cylindrical lens
attached to a micromachined oscillator and interacting
with a plate via the Casimir force in the dynamic
regime. We derive the
exact expression for a shift of the natural frequency of
the oscillator under the influence of the Casimir force.
This allows measurements of the frequency shift in a
nonlinear regime, and comparison of the
experimental results
with different theoretical approaches to the Casimir
force.

The paper is organized as follows. In Sec.~II the
experimental procedures for microfabrication of smooth
cylindrical objects are considered. In Sec.~III the
Lifshitz-type formulas are derived for the Casimir force
and for the gradient of the Casimir force between a
plate and an elliptic cylinder. In Sec.~IV the same
is done for an asymmetric cylindrical lens near a
plate. Section~V is devoted to the micromachined
oscillator with an attached cylindrical lens under
the influence of the Casimir force in a nonlinear
regime. Section~VI contains our conclusions and
discussion.

\section{Techniques for microfabrication of smooth cylindrical
objects}

There are several approaches to creating a smooth
cylindrical object on  top of a micromachined torsional
oscillator that are fully compatible with
ion chromatography
techniques. As a consequence, the cylindrical microstructures
can be monolithically integrated with the
microelectromechanical oscillators.
Examples include
Focused Ion Beam technology (FIB) and
techniques based on femtosecond laser microfabrication.

FIB technology uses a Ga${}^+$ ion beam to
remove material from almost any surface. The profile to be
patterned can be automatically inputted and
controlled rather precisely.
These tools are available in almost any microfabrication
laboratory and, when combined with a scanning electron
microscope, they can be used for
non-destructive imaging at higher magnifications, permitting
extremely accurate control of the milling process
\cite{T1,T2}. Today's most advanced FIB tools allow direct
patterning of metals with minimum contamination or damage,
 which opens up the possibility to directly pattern nanostructures
with desirable shapes onto microelectromechanical oscillators.

Techniques based on femtosecond laser microfabrication, similar
to the ones used to fabricate microlenses on glass \cite{T3}, can
be also used to integrate a cylinder onto the paddle of a
microelectromechanical system.
In this case a tightly focused femtosecond laser beam is scanned
inside a photosensitive material to create the required
shape as precisely as possible.
Once the photosensitive material is developed, the exposed
volume will remain on the oscillator plate and standard etching
processes can be used to transfer this shape to the plate.

These microfabrication techniques represent, in our opinion,
the most versatile methods of fabricating microstructures
of a desirable shape.
For objects of cylindrical shape, microfabrication typically
results in elliptic rather than circular cylinders.
The actual shape of a microfabricated object can be measured
very precisely  using a noncontact optical profilometer.
Microfabricated cylinders may have semiaxes in two
perpendicular directions varying by 20\%--30\%.
They might be also characterized by some asymmetry (for
instance, the axis of a cylinder may be not exactly parallel
to the plate of a microelectromechanical oscillator).
More complex and expensive fabrication methods can also be used depending
on the precision, uniformity and reproducibility required \cite{T4,T5}.

\section{The Casimir force between an elliptic cylinder and
a plate within the proximity force approximation}

We consider an elliptic cylindrical lens of thickness $h$
and width $2d$ obtained from an elliptic cylinder made of a
material with a frequency-dependent dielectric permittivity
$\varepsilon(\omega)$. Let the surface of this cylinder be
described by the equation
\begin{equation}
\frac{x^2}{A^2}+\frac{(z-a-B)^2}{B^2}=1,
\label{eq1}
\end{equation}
\noindent
where $A>B$ are semiaxes and $a$ is the closest separation
distance between the lens and the plate $z=0$ (see Fig.~1).
The axis of a cylinder is aligned along the $y$ axis.
The upper surface of a plate made of the same material
coincides with the plane $z=0$.
The elliptic lens under consideration is assumed to be
attached to a micromachined oscillator (see Fig.~4 in
Ref.~\cite{33} where a circular cylindrical lens is
situated {\it below} a plate). In Fig.~1 the plate is
placed below a cylindrical lens for convenience in
calculations.

{}From Eq.~(\ref{eq1}) the explicit equation for the
lens surface is given by
\begin{equation}
z(x)=a+B-\sqrt{B^2-\frac{B^2}{A^2}x^2}.
\label{eq2}
\end{equation}
\noindent
It is assumed that $a/B\ll 1$ and the lens is sufficiently thick,
so that $a/h\ll 1$ as well.
Applying the PFA in the general, Derjaguin, formulation
\cite{2,34} to the configuration of Fig.~1 in the same way as
was done in Ref.~\cite{33} for a circular cylinder, one arrives
to the following Lifshitz-type formula for the Casimir force
at temperature $T$:
\begin{eqnarray}
&&
F(a,T)=-\frac{2k_BTL}{\pi}
\sum_{l=0}^{\infty}{\vphantom{\sum}}^{\!\prime}
\int_0^{\infty}q_lk_{\bot}dk_{\bot}
\nonumber \\
&&~~~~~~~~~
\times
\sum_{n=1}^{\infty}(r_{\rm TM}^{2n}+r_{\rm TE}^{2n})
\int_{0}^{d}dxe^{-2nq_lz(x)}.
\label{eq3}
\end{eqnarray}
\noindent
Here, $k_B$ is the Boltzmann constant,
$L$ is the length of the cylinder which is assumed to be
infinitely large,
$k_{\bot}$ is the
projection of the wave vector on the plane $z=0$,
$q_l=(k_{\bot}^2+\xi_l^2/c^2)^{1/2}$,
$\xi_l=2\pi k_B Tl/\hbar$ with $l=0,\,1,\,2,\,\ldots$ are the
Matsubara frequencies, and $z(x)$ is defined in Eq.~(\ref{eq2}).
 The primed summation
means that the term with $l=0$ is multiplied by 1/2.
The reflection coefficients $r_{\rm TM}$  and $r_{\rm TE}$ for the
two polarizations of the electromagnetic field (transverse magnetic
 and transverse electric) are given by
\begin{eqnarray}
&&
r_{\rm TM}=r_{\rm TM}(i\xi_l,k_{\bot})=
\frac{\varepsilon_lq_l-k_l}{\varepsilon_lq_l+k_l},
\nonumber \\
&&
r_{\rm TE}=r_{\rm TE}(i\xi_l,k_{\bot})=
\frac{q_l-k_l}{q_l+k_l},
\label{eq4}
\end{eqnarray}
\noindent
where
$k_l=\left[k_{\bot}^2+\varepsilon_l{\xi_l^2}/{c^2}\right]^{1/2}$
and $\varepsilon_l=\varepsilon(i\xi_l)$.

Bearing in mind that in accordance with Eq.~(\ref{eq2})
\begin{equation}
x=x(z)=\frac{A}{B}\sqrt{B^2-(a+B-z)^2},
\label{eq5}
\end{equation}
\noindent
one can rearrange Eq.~(\ref{eq3}) to the form
\begin{eqnarray}
&&
F(a,T)=-\frac{2k_BTL}{\pi}
\sum_{l=0}^{\infty}{\vphantom{\sum}}^{\!\prime}
\int_0^{\infty}q_lk_{\bot}dk_{\bot}
\nonumber \\
&&~~~~~~~~~~
\times\sum_{n=1}^{\infty}(r_{\rm TM}^{2n}+r_{\rm TE}^{2n})
\int_{a}^{a+h}\!\!\!dx(z)e^{-2nq_lz}
\nonumber \\
&&
=-\frac{2k_BTLA}{\pi B}
\sum_{l=0}^{\infty}{\vphantom{\sum}}^{\!\prime}
\int_0^{\infty}q_lk_{\bot}dk_{\bot}
\label{eq6}\\
&&~~
\times
\sum_{n=1}^{\infty}(r_{\rm TM}^{2n}+r_{\rm TE}^{2n})
\int_{a}^{a+h}\!\!\!\frac{(a+B-z)e^{-2nq_lz}}{\sqrt{B^2-(a+B-z)^2}}\,dz.
\nonumber
\end{eqnarray}
\noindent
We next introduce dimensionless integration variables
\begin{equation}
v=2aq_l, \qquad t=nv\frac{z-a}{a}
\label{eq7}
\end{equation}
\noindent
instead of dimensional $k_{\bot}$ and $z$, and dimensionless
Matsubara frequencies $\zeta_l=2a\xi_l/c$. As a result, from
Eq.~(\ref{eq6}) we arrive at the expression
\begin{eqnarray}
&&
F(a,T)=-\frac{k_BTLA}{4\pi a^2 B}
\sum_{l=0}^{\infty}{\vphantom{\sum}}^{\!\prime}
\sum_{n=1}^{\infty}\frac{1}{n}\int_{\zeta_l}^{\infty}vdv
(r_{\rm TM}^{2n}+r_{\rm TE}^{2n})
\nonumber\\
&&~~
\times
e^{-nv}\int_{0}^{hnv/a}\!\!
\frac{1-\frac{a}{Bnv}t}{\sqrt{1-(1-\frac{a}{Bnv}t)^2}}\,
e^{-t}dt.
\label{eq8}
\end{eqnarray}
\noindent
Here, the reflection coefficients in terms of new variables are
given by
\begin{eqnarray}
&&
r_{\rm TM}=r_{\rm TM}(i\zeta_l,v)=
\frac{\varepsilon_lv-\sqrt{v^2+(\varepsilon_l-1)\zeta_l^2}}{\varepsilon_lv+
\sqrt{v^2+(\varepsilon_l-1)\zeta_l^2}},
\nonumber \\
&&
r_{\rm TE}=r_{\rm TE}(i\zeta_l,v)=
\frac{v-\sqrt{v^2+(\varepsilon_l-1)\zeta_l^2}}{v+
\sqrt{v^2+(\varepsilon_l-1)\zeta_l^2}},
\label{eq9}
\end{eqnarray}
\noindent
where $\varepsilon_l=\varepsilon(ic\zeta_l/2a)$.

Within the application regime of the PFA we are looking for
the main contribution to the expansion of Eq.~(\ref{eq8}) in
terms of the
small parameter $a/B\ll 1$. To do so we can restrict our
consideration to the lowest expansion order in $a/B$ of the
integrand with respect to $t$. We can also set the upper
integration limit of this integral equal to infinity
taking into account that $h\gg a$. This leads to
\begin{eqnarray}
&&
F(a,T)=-\frac{k_BTL}{4\pi a^2}\,\frac{A}{\sqrt{2aB}}
\sum_{l=0}^{\infty}{\vphantom{\sum}}^{\!\prime}
\sum_{n=1}^{\infty}\frac{1}{\sqrt{n}}
\label{eq10}\\
&&~~
\times\int_{\zeta_l}^{\infty}v^{3/2}dv
(r_{\rm TM}^{2n}+r_{\rm TE}^{2n})
e^{-nv}\int_{0}^{\infty}\!\!
\frac{e^{-t}dt}{\sqrt{t}}.
\nonumber
\end{eqnarray}
\noindent
After calculation of an integral with respect to $t$, and
summation with respect to $n$ one obtains
\begin{eqnarray}
&&
F(a,T)=-\frac{k_BTL}{4\sqrt{\pi} a^2}\,\frac{A}{\sqrt{2aB}}
\sum_{l=0}^{\infty}{\vphantom{\sum}}^{\!\prime}
\int_{\zeta_l}^{\infty}v^{3/2}dv
\nonumber\\
&&~~
\times\left[
{\rm Li}_{1/2}(r_{\rm TM}^{2}e^{-v})+
{\rm Li}_{1/2}(r_{\rm TE}^{2}e^{-v})\right],
\label{eq11}
\end{eqnarray}
\noindent
where ${\rm Li}_n(z)$ is the polylogarithm function
\cite{PBM}.
This is the Lifshitz-type formula for the Casimir
force between an elliptic cylinder or cylindrical lens and
a plate. For a circular cylinder $A=B=R$, and Eq.~(\ref{eq11})
coincides with the result derived in Ref.~\cite{33}.

In the case of an elliptic cylinder  and a plate made of an
ideal metal, $r_{\rm TM}^2=r_{\rm TE}^2=1$. Then at zero
temperature Eq.~(\ref{eq11}) results in
\begin{eqnarray}
&&
F^{\rm IM}(a,0)=-\frac{L\hbar c}{8\pi\sqrt{\pi} a^3}\,\frac{A}{\sqrt{2aB}}
\int_{0}^{\infty}d\zeta
\nonumber \\
&&~~~~~~
\times
\int_{\zeta}^{\infty}v^{3/2}dv
\sum_{n=1}^{\infty}\frac{e^{-nv}}{\sqrt{n}}.
\label{eq12}
\end{eqnarray}
\noindent
Changing the order of integrations and calculating the integrals
one obtains
\begin{equation}
F^{\rm IM}(a,0)=-\frac{15L\hbar c}{64\pi a^3}\,\frac{A}{\sqrt{2aB}}
\sum_{n=1}^{\infty}\frac{1}{n^4}.
\label{eq13}
\end{equation}
\noindent
After calculating the sum, the result is
\begin{equation}
F^{\rm IM}(a,0)=-\frac{\pi^3L\hbar c}{384 a^3}\,\frac{A}{\sqrt{2aB}}.
\label{eq14}
\end{equation}
\noindent
 For a circular cylinder ($A=B=R$) this leads to a well known
result obtained in Ref.~\cite{30}.

Both Eqs.~(\ref{eq11}) and (\ref{eq14}) are approximate,
as they are obtained with the help of the PFA. Using the same
considerations as in Ref.~\cite{33}, one can conclude that the
relative error of these equations is approximately $0.3a/B$.
For typical experimental parameters $B=100\,\mu$m and $a=200\,$nm
the resulting error is equal to 0.06\%.

The Lifshitz-type formula for the gradient of the Casimir force
between an elliptic cylinder and a plate can be obtained by
analogy with Eq.~(\ref{eq11}). For this purpose we differentiate
Eq.~(\ref{eq3}) with respect to $a$ using Eq.~(\ref{eq2})
and arrive at
\begin{eqnarray}
&&
\frac{\partial F(a,T)}{\partial a}
=\frac{4k_BTL}{\pi}
\sum_{l=0}^{\infty}{\vphantom{\sum}}^{\!\prime}
\int_0^{\infty}q_l^2k_{\bot}dk_{\bot}
\nonumber \\
&&~~~~~~~~~
\times
\sum_{n=1}^{\infty}n(r_{\rm TM}^{2n}+r_{\rm TE}^{2n})
\int_{0}^{d}dxe^{-2nq_lz(x)}.
\label{eq15}
\end{eqnarray}
\noindent
We then repeat the same transformations as were done in
Eqs.~(\ref{eq6})--(\ref{eq10}) in application to Eq.~(\ref{eq15})
and obtain
\begin{eqnarray}
&&
\frac{\partial F(a,T)}{\partial a}
=\frac{k_BTL}{4\sqrt{\pi} a^3}\,\frac{A}{\sqrt{2aB}}
\sum_{l=0}^{\infty}{\vphantom{\sum}}^{\!\prime}
\int_{\zeta_l}^{\infty}v^{5/2}dv
\nonumber\\
&&~~
\times\left[
{\rm Li}_{-1/2}(r_{\rm TM}^{2}e^{-v})+
{\rm Li}_{-1/2}(r_{\rm TE}^{2}e^{-v})\right].
\label{eq16}
\end{eqnarray}

For a circular cylinder, Eq.~(\ref{eq16}) coincides with
the result
obtained in Ref.~\cite{33}. In the case of an ideal metal
elliptic cylinder and a plate at zero temperature
Eq.~(\ref{eq16}) leads to
\begin{equation}
\frac{\partial F^{\rm IM}(a,0)}{\partial a}=
\frac{7\pi^3L\hbar c}{768 a^3}\,\frac{A}{\sqrt{2aB}}.
\label{eq17}
\end{equation}
\noindent
The same equation is obtained by differentiation of
Eq.~(\ref{eq14}) with respect to $a$.

The Lifshitz-type formulas (\ref{eq11}) and (\ref{eq16})
allow computations of the Casimir force and its gradient
in the configuration of an elliptic cylindrical lens and
a plate made of real materials. In so doing different
theoretical approaches can be used, such as the Drude
and plasma model approaches mentioned in Sec.~I.
For an Au circular cylinder above a plate computations
of the relative thermal correction to the Casimir force and its
gradient as a function of separation using the Drude and
plasma model approaches were performed in Ref.~\cite{33}
within the separation range from 150\,nm to $5\,\mu$m.
It was shown that when the Drude model is used the
magnitude of the relative thermal corection to the
Casimir force achieves its maximum value 41.6\% at
$a=2.55\,\mu$m. The maximum magnitude of the
relative thermal
correction to the gradient of the Casimir force 52\%
occurs at $a=3.6\,\mu$m. When the plasma model approach
is used, the relative thermal correction to the Casimir
force increases monotonically from 0.016\% at 150\,nm
to 26.7\% at $a=5\,\mu$m \cite{33}.
In the case of an elliptic cylinder the respective
results for the relative thermal correction remain the
same. This allows discrimination between the predictions of
different theoretical approaches by comparing the
computation results with the measurement data.

By using the PFA, one can also obtain a simple expression for
the electric force between an elliptic cylinder and a
plate. An electric force is used to perform calibrations in
the measurements of the Casimir force. For a potential
difference $(V-V_0)$ between an elliptic cylinder and a
plate ($V$ is the applied voltage and $V_0$ is the
residual potential), the electric force calculated
similar to the Casimir force is given by
\begin{equation}
F_{\rm el}(a)=-\frac{\pi\epsilon_0L}{2a}\,
\frac{A}{\sqrt{2aB}}(V-V_0)^2,
\label{eq17a}
\end{equation}
\noindent
where $\epsilon_0$ is the permittivity of the vacuum.
For a circular cylinder $A=B=R$, this formula was
obtained \cite{31} from the exact expression for the
electric force \cite{35a}
\begin{equation}
F_{\rm el}(a)=\frac{4\pi\epsilon_0L(V-V_0)^2}{\Delta
\ln^2\left(\frac{h-\Delta}{h+\Delta}\right)},
\label{eq18a}
\end{equation}
\noindent
where $\Delta=\sqrt{h^2-R^2}$ and $h=R+a$.
Expanding the right-hand side of Eq.~(\ref{eq18a}) in
powers of a small parameter $a/R$, one obtains
\begin{equation}
F_{\rm el}(a)=-\frac{\pi\epsilon_0L\sqrt{R}}{2\sqrt{2}a^{3/2}}
(V-V_0)^2\left(1-\frac{1}{12}\,\frac{a}{R}+
\frac{17}{480}\,\frac{a^2}{R^2}\right).
\label{eq18b}
\end{equation}
\noindent
Thus, for $a=100\,$nm and $R=100\,\mu$m the error in the
electric force due to the use of the PFA is equal
to only 0.008\%.

\section{An asymmetric cylindrical lens and a plate}

The above results can be used to calculate the Casimir
force between an asymmetric cylindrical lens modeled
by the two elliptic cylinders with dissimilar semiaxes
$A_1,\,B_1$ and $A_2,\,B_2$ (see Fig.~2).
One half of such a lens of width $d_1$ is produced as
a section of an elliptic cylinder with semiaxes
$A_1,\,B_1$,
and another half of width $d_2$  as
a section of a cylinder with semiaxes $A_2,\,B_2$.
In so doing both halves are equal in thickness.
Bearing in mind that the PFA is an additive method,
the Casimir force between each of the halves of an
asymmetric cylindrical lens and a plate can be
calculated using Eq.~(\ref{eq11}). The Casimir
force between the entire lens and a plate is
then given by
\begin{eqnarray}
&&
F(a,T)=-\frac{k_BTL}{8\sqrt{\pi} a^2}\,\frac{1}{\sqrt{2a}}
\left(\frac{A_1}{\sqrt{B_1}}+\frac{A_2}{\sqrt{B_2}}\right)
\label{eq18}\\
&&~~
\times
\sum_{l=0}^{\infty}{\vphantom{\sum}}^{\!\prime}
\int_{\zeta_l}^{\infty}\!\!\!v^{3/2}dv
\left[
{\rm Li}_{1/2}(r_{\rm TM}^{2}e^{-v})+
{\rm Li}_{1/2}(r_{\rm TE}^{2}e^{-v})\right].
\nonumber
\end{eqnarray}
\noindent
This equation is valid under the conditions $a/B_1\ll 1$,
$a/B_2\ll 1$, and $a/h\ll 1$.
In a similar way, the gradient of the Casimir force
between an asymmetric elliptic lens shown in Fig.~2 and
a plate is expressed by the equation
\begin{eqnarray}
&&
\frac{\partial F(a,T}{\partial a})=
\frac{k_BTL}{8\sqrt{\pi} a^3}\,\frac{1}{\sqrt{2a}}
\left(\frac{A_1}{\sqrt{B_1}}+\frac{A_2}{\sqrt{B_2}}\right)
\label{eq19}\\
&&
\times
\sum_{l=0}^{\infty}{\vphantom{\sum}}^{\!\prime}
\int_{\zeta_l}^{\infty}\!\!\!v^{5/2}dv
\left[
{\rm Li}_{-1/2}(r_{\rm TM}^{2}e^{-v})+
{\rm Li}_{-1/2}(r_{\rm TE}^{2}e^{-v})\right].
\nonumber
\end{eqnarray}
\noindent
Equations (\ref{eq18}) and  (\ref{eq19}) allow calculation of
the Casimir force and its gradient in the configuration of a
plate and a cylindrical lens consisting of two parts of
dissimilar elliptic cylinders.

We turn next to the consideration of another asymmetric
cylindrical lens which is obtained from an elliptic cylinder
defined in its proper coordinates ($\tilde{x},\tilde{z}$)
as a cross section by the plane perpendicular to the
plane $\tilde{x}\tilde{z}$ and inclined at an angle
$\varphi$ to the axis $\tilde{x}$ [see Fig.~3(a)].
We then rotate the resulting lens through an angle
$\varphi$ clockwise around the axis of a cylinder in order to
make its base parallel to the plate [see Fig.~3(b)].
As before, the thickness of a lens is $h$.

It is easily seen that the transformation from the
coordinates ($x,z$) to ($\tilde{x},\tilde{z}$) shown in
Fig.~3(b) has the form
\begin{eqnarray}
&&
\tilde{x}=\tilde{x}_0+a\sin\varphi+x\cos\varphi-z\sin\varphi,
\nonumber \\
&&
\tilde{z}=\tilde{z}_0-a\cos\varphi+x\sin\varphi+z\cos\varphi,
\label{eq20}
\end{eqnarray}
\noindent
where ($\tilde{x}_0,\tilde{z}_0$) are the coordinates of the lens
point closest to the plate given by
\begin{equation}
\tilde{x}_0=\frac{A^2\sin\varphi }{H},
\quad
\tilde{z}_0=-\frac{B^2\cos\varphi}{H}.
\label{eq21}
\end{equation}
\noindent
Here, we have introduced the notation
\begin{equation}
H\equiv H(A,B;\varphi)=\sqrt{A^2\sin^2\varphi+
B^2\cos^2\varphi}.
\label{eq21a}
\end{equation}
\noindent
Now we substitute Eqs.~(\ref{eq20}) and (\ref{eq21}) into the
equation of a lens surface
\begin{equation}
\frac{\tilde{x}^2}{A^2}+\frac{\tilde{z}^2}{B^2}=1
\label{eq22}
\end{equation}
\noindent
written in the proper coordinates and arrive at
\begin{eqnarray}
&&
x^2-2x\frac{A^2-B^2}{H^2}(a-z)\sin\varphi\cos\varphi+
2(a-z)\frac{A^2B^2}{H^3}
\nonumber \\
&&~~~~~~~~~
+(a-z)^2\,
\frac{A^2\cos^2\varphi+B^2\sin^2\varphi}{H^2}=0.
\label{eq23}
\end{eqnarray}
\noindent
This equation describes the surface of an asymmetric cylindrical
lens in the coordinates ($x,z$). If the inclination angle is
$\varphi=0$, Eq.~(\ref{eq23}) simplifies to
\begin{equation}
x^2+2(a-z)\frac{A^2}{B}+
(a-z)^2\frac{A^2}{B^2}=0,
\label{eq24}
\end{equation}
\noindent
and has the solution (\ref{eq5}) as it must. For a circular
cylinder $A=B=R$ and an arbitrary angle $\varphi$,
 Eq.~(\ref{eq23}) simplifies to
\begin{equation}
x^2+2(a-z)R+
(a-z)^2=0,
\label{eq25}
\end{equation}
\noindent
leading again to the specific case of Eq.~(\ref{eq5}).

Equation (\ref{eq23}) has the following two solutions:
\begin{eqnarray}
&&
x_{1,2}=-\frac{A^2-B^2}{H^2}\,(z-a)\sin\varphi\cos\varphi
\nonumber \\
&&~~~
\pm\frac{AB}{H^2}\left[2(z-a)H-(z-a)^2\right]^{1/2},
\label{eq26}
\end{eqnarray}
\noindent
where the upper and lower signs are for $x>0$ and $x<0$,
respectively [see Fig.~3(b)].

We next consider the calculation of the thermal Casimir force
between an asymmetric cylindrical lens and a plate shown in
Fig.~3(b). This can be done by using the first equality in
Eq.~(\ref{eq6}) which we apply separately to the parts of the
lens with $x<0$ and $x>0$:
\begin{eqnarray}
&&
F_{\varphi}(a,T)=-\frac{k_BTL}{\pi}
\sum_{l=0}^{\infty}{\vphantom{\sum}}^{\!\prime}
\int_0^{\infty}\!\!\!q_lk_{\bot}dk_{\bot}
\sum_{n=1}^{\infty}(r_{\rm TM}^{2n}+r_{\rm TE}^{2n})
\nonumber \\
&&~~~~~
\times
\left[\int_{a+h}^{a}\!\!\!dx_2(z)e^{-2nq_lz}+
\int_{a}^{a+h}\!\!\!dx_1(z)e^{-2nq_lz}\right].
\label{eq27}
\end{eqnarray}
\noindent
{}From Eq.~(\ref{eq26}), the differentials $dx_{1,2}$
are given by
\begin{eqnarray}
&&
dx_{1,2}=-\frac{(A^2-B^2)\sin\varphi\cos\varphi}{H^2}dz
\nonumber \\
&&~~~
\pm\frac{AB}{H^2}
\frac{H-z+a}{\left[2(z-a)H-(z-a)^2\right]^{1/2}}dz
\label{eq28}
\end{eqnarray}
\noindent
with the same sign convention as formulated above.
Substituting Eq.~(\ref{eq28}) into Eq.~(\ref{eq27}),
one obtains
\begin{eqnarray}
&&
F_{\varphi}(a,T)=-\frac{2k_BTL}{\pi}\,\frac{AB}{H^2}
\sum_{l=0}^{\infty}{\vphantom{\sum}}^{\!\prime}
\int_0^{\infty}\!\!\!q_lk_{\bot}dk_{\bot}
\label{eq29} \\
&&~
\times
\sum_{n=1}^{\infty}(r_{\rm TM}^{2n}+r_{\rm TE}^{2n})
\int_{a}^{a+h}\!\!\!
\frac{(H-z+a)e^{-2nq_lz}dz}{\left[2(z-a)H-(z-a)^2\right]^{1/2}}.
\nonumber
\end{eqnarray}
\noindent
Introducing the integration variable $v$ from Eq.~(\ref{eq7})
instead of the variable $k_{\bot}$, this can be rearranged to
\begin{eqnarray}
&&
F_{\varphi}(a,T)=-\frac{k_BTL}{4\pi a^3}\,\frac{AB}{H^2}
\sum_{l=0}^{\infty}{\vphantom{\sum}}^{\!\prime}
\int_{\zeta_l}^{\infty}\!\!\!v^2dv
\label{eq30} \\
&&~
\times
\sum_{n=1}^{\infty}(r_{\rm TM}^{2n}+r_{\rm TE}^{2n})
\int_{a}^{a+h}\!\!\!
\frac{(H-z+a)e^{-nvz/a}dz}{\left[2(z-a)H-(z-a)^2\right]^{1/2}}.
\nonumber
\end{eqnarray}
\noindent
Now, instead of the variable $z$, we introduce the variable $t$
defined in Eq.~(\ref{eq7}) and use the conditions
\begin{equation}
\frac{a}{h}\ll 1, \qquad
\frac{a}{H}\ll 1.
\label{eq31}
\end{equation}
\noindent
Then Eq.~(\ref{eq30}) reduces to
\begin{eqnarray}
&&
F_{\varphi}(a,T)=-\frac{k_BTL}{4\sqrt{2}\pi a^{5/2}}\,\frac{AB}{H^{3/2}}
\sum_{l=0}^{\infty}{\vphantom{\sum}}^{\!\prime}
\int_{\zeta_l}^{\infty}\!\!\!v^{3/2}dv
\nonumber \\
&&~
\times
\sum_{n=1}^{\infty}\frac{1}{\sqrt{n}}(r_{\rm TM}^{2n}+r_{\rm TE}^{2n})
e^{-nv}
\int_{0}^{\infty}\!\!\!
\frac{e^{t}}{\sqrt{t}}dt.
\label{eq32}
\end{eqnarray}
\noindent
Performing the integration with respect to $t$ and the summation
over $n$, we arrive at
\begin{eqnarray}
&&
F_{\varphi}(a,T)=-\frac{k_BTL}{4\sqrt{\pi} a^2}\,\frac{A}{\sqrt{2aB}}
\left(\frac{B}{H}\right)^{3/2}
\sum_{l=0}^{\infty}{\vphantom{\sum}}^{\!\prime}
\int_{\zeta_l}^{\infty}\!\!\!v^{3/2}dv
\nonumber\\
&&
\times
\left[
{\rm Li}_{1/2}(r_{\rm TM}^{2}e^{-v})+
{\rm Li}_{1/2}(r_{\rm TE}^{2}e^{-v})\right].
\label{eq33}
\end{eqnarray}
\noindent
The comparison of this result with Eq.~(\ref{eq11}) shows that the
dependence of the Casimir force on $\varphi$ is contained
exclusively in the factor
\begin{equation}
G(A,B;\varphi)=\left(\frac{B}{H}\right)^{3/2}\!\!\!=
\left(\frac{A^2}{B^2}\sin^2\varphi+\cos^2\varphi\right)^{-{3/4}}.
\label{eq34}
\end{equation}
\noindent
Thus, a similar  result is obtained for the gradient of the
Casimir force between an asymmetric cylindrical lens and a plate
\begin{equation}
\frac{\partial F_{\varphi}(a,T)}{\partial a}=
G(A,B;\varphi)\frac{\partial F(a,T)}{\partial a},
\label{eq35}
\end{equation}
\noindent
where the gradient of the Casimir force between a symmetric
elliptic cylindrical lens and a plate,
$\partial F(a,T)/\partial a$, is given by Eq.~(\ref{eq16}).

{}From Eq.~(\ref{eq34}) it is seen that the function $G$ and,
thus, the Casimir force $F_{\varphi}$ and its gradient satisfy
the condition
\begin{equation}
\frac{A}{\sqrt{B}}G\left(A,B;\varphi+\frac{\pi}{2}\right)=
\frac{B}{\sqrt{A}}G(B,A;\varphi).
\label{eq36}
\end{equation}
\noindent
Specifically, from Eq.~(\ref{eq36}) we have
\begin{equation}
\frac{A}{\sqrt{B}}G\left(A,B;\frac{\pi}{2}\right)=
\frac{B}{\sqrt{A}},
\label{eq37}
\end{equation}
\noindent
i.e., the rotation through an angle $\varphi=\pi/2$ interchanges the
semiaxes of a cylinder, as it should.

In Fig.~4(a) we present the relative Casimir force and its
gradient
\begin{equation}
\frac{F_{\varphi}(a,T)}{F(a,T)}=
\frac{\partial F_{\varphi}(a,T)/\partial a}{\partial F(a,T)/\partial a}
=G(A,B;\varphi)
\label{eq38}
\end{equation}
\noindent
as a function of the rotation angle. Different lines are for
different values of the ratio of semiaxes $A/B=1.1$, 1.2, 1.3,
and 1.4 increasing from the top to bottom lines.
Keeping in mind that experimentally it is difficult to ensure
exactly $\varphi=0$, we also present  in Fig.~4(b) the same lines
over a narrow interval from $\varphi=0$ to $\varphi=0.1\,$rad.
{}From Fig.~4(a) it is seen that the relative Casimir force and
its gradient decrease monotonically with the increase of
$\varphi$ and $A/B$. According to Fig.~4(b), even the rotation
of an elliptic lens through 0.1\,rad ($5.73^{\circ}$) leads to less
than  1\% deviation of the Casimir force and its gradient from
their respective values at $\varphi=0$ for any value of $A/B$
considered. A deviation of the Casimir force and its gradient from
their values at $\varphi=0$ for less than 0.1\% is achieved for the
rotation angles $\varphi<0.025\,\mbox{rad}=1.43^{\circ}$.
This places experimental limitations on an allowed
asymmetry of the elliptic cylindrical lens used.

Note that the electric force between an asymmetric cylinder and
a plate can be obtained from Eq.~(\ref{eq17a}) by the replacement
\begin{equation}
\frac{A}{\sqrt{B}}\to\frac{1}{2}\left(
\frac{A_1}{\sqrt{B_1}}+\frac{A_2}{\sqrt{B_2}}\right),
\label{eq39}
\end{equation}
\noindent
or by multiplying the right-hand side of Eq.~(\ref{eq17a}) by
the factor $G$ defined in Eq.~(\ref{eq34}) depending on the
nature of asymmetry.

\section{Micromachined oscillator with an attached cylindrical
lens in a nonlinear regime}

In Ref.~\cite{33} it was proposed to perform dynamic measurements
of the Casimir interaction between a plate and a circular cylinder
attached to a micromachined oscillator.
The proposed experiment aims to achieve the same high experimental
precision, as in the experiment of Refs.~\cite{15,16} for a
sphere above a plate, over a wider separation region.
For this purpose, the same measures, as in Refs.~\cite{15,16}
would
be undertaken, specifically, to reduce mechanical vibrations.
At any rate, the effect of vibrations in the position at the
proposed measurement frequency (a few hundred hertz) is much
smaller than the uncertainty in the position due to the
interferometric technique used. As a result, the impact of
vibrations on the determination of the gradient of the Casimir
force is smaller than the current systematic experimental error,
and thus, can be neglected.

As in Refs.~\cite{15,16,16a,37,38} which exploited the configuration of a sphere
near a plate, Ref.~\cite{33} discussed measurements of the
gradient of the Casimir force in a linear regime (the same regime
was employed in dynamic measurements by means of an atomic force
microscope \cite{39,40,41,42,43}). Here, we find the frequency
shift of an oscillator, caused by the Casimir force between an
elliptic cylinder and a plate, in the nonlinear regime.
This allows measurements down to shorter separation distances
where the micromachined oscillator behaves nonlinearly.

In the dynamic regime the separation distance between an elliptic
cylinder attached to a micromachined oscillator and a plate is
varied with time harmonically
\begin{equation}
a(t)=a+A_z\cos\omega_rt.
\label{eqN1}
\end{equation}
\noindent
Here, $\omega_r$ is the resonant frequency of the oscillator
under the influence of the Casimir force acting between a
cylinder and a plate. The amplitude of oscillations $A_z$
should be sufficiently small in comparison with the
separation $a$. In the presence of the Casimir force,
$F(a,T)$, the frequency $\omega_r$ is different from the
natural angular frequency of the oscillator $\omega_0$.
Such an oscillator problem was considered in
Refs.~\cite{35,36} perturbatively and in Ref.~\cite{44}
exactly. The exact expression for the shift of the
second power of the
natural frequency of an oscillator produced by
the Casimir force is given by \cite{44}
\begin{equation}
\omega_r^2-\omega_0^2=-\frac{C\omega_r}{\pi A_z}
\int_{0}^{2\pi/\omega_r}\!\!\!\!dt
\cos(\omega_rt)F(a+A_z\cos\omega_rt,T).
\label{eqN2}
\end{equation}
\noindent
Here, $C$ is a constant depending on specific parameters of
the setup used. Thus, for a micromachined oscillator
$C=b^2/I$, where $b$ and $I$ are the lever arm and the moment
of inertia. Note that in Ref.~\cite{44}, where the
Bose-Einstein condensate was considered as a second body,
the Casimir-Polder force between individual atoms and a
plate was also averaged over the condensate cloud.

In the case of an elliptic cylinder interacting with the plate
the force $F$ is given by Eq.~(\ref{eq11}). Representing the
polilogarithm functions in Eq.~(\ref{eq11}) as power series,
and replacing the integration variable $v$ with
$k_{\bot}=\sqrt{v^2-\zeta_l^2}/(2a)$, we rearrange this
equation to the form
\begin{eqnarray}
&&
F(a,T)=-\frac{k_BTL}{\sqrt{\pi}}\,\frac{A}{\sqrt{B}}
\sum_{l=0}^{\infty}{\vphantom{\sum}}^{\!\prime}
\sum_{n=1}^{\infty}\frac{1}{\sqrt{n}}
\int_{0}^{\infty}\!\!\!\!k_{\bot}dk_{\bot}\sqrt{q_l}
\nonumber\\
&&~~
\times
(r_{\rm TM}^{2n}+r_{\rm TE}^{2n})
e^{-2aq_ln}.
\label{eqN3}
\end{eqnarray}
\noindent
Substituting Eq.~(\ref{eqN3}) into Eq.~(\ref{eqN2}) and
introducing the new integration variable $\theta=\omega_rt$,
we arrive at
\begin{eqnarray}
&&
\omega_r^2-\omega_0^2=\frac{C}{\pi A_z}\frac{k_BTL}{\sqrt{\pi}}\,\frac{A}{\sqrt{B}}
\sum_{l=0}^{\infty}{\vphantom{\sum}}^{\!\prime}
\sum_{n=1}^{\infty}\frac{1}{\sqrt{n}}
\int_{0}^{\infty}\!\!\!\!k_{\bot}dk_{\bot}\sqrt{q_l}
\nonumber\\
&&~
\times
(r_{\rm TM}^{2n}+r_{\rm TE}^{2n})
e^{-2aq_ln}\!
\int_{0}^{2\pi}\!d\theta\cos\theta e^{-2nA_zq_l\cos\theta}.
\label{eqN4}
\end{eqnarray}
\noindent
Integrating by parts, the integral with respect to $\theta$
can be reduced to \cite{45}
\begin{equation}
\int_{0}^{\pi}\!\sin\theta e^{\pm z\cos\theta}d\theta=
\frac{\pi}{2}\,{I}_1(z),
\label{eqN5}
\end{equation}
\noindent
where ${I}_k(z)$ is the Bessel function of an imaginary
argument. The frequency shift (\ref{eqN4}) is then expressed
as
\begin{eqnarray}
&&
\omega_r^2-\omega_0^2=-\frac{2C}{ A_z}\frac{k_BTL}{\sqrt{\pi}}\,\frac{A}{\sqrt{B}}
\sum_{l=0}^{\infty}{\vphantom{\sum}}^{\!\prime}
\sum_{n=1}^{\infty}\frac{1}{\sqrt{n}}
\int_{0}^{\infty}\!\!\!\!k_{\bot}dk_{\bot}
\nonumber\\
&&~
\times
\sqrt{q_l}(r_{\rm TM}^{2n}+r_{\rm TE}^{2n})
e^{-2aq_ln}{I}_1(2A_zq_ln).
\label{eqN6}
\end{eqnarray}
\noindent
Returning to the variable $v$, one finally obtains
\begin{eqnarray}
&&
\omega_r^2-\omega_0^2=-\frac{C}{ A_z}\frac{k_BTL}{2\sqrt{\pi}a^2}\,\frac{A}{\sqrt{2aB}}
\sum_{l=0}^{\infty}{\vphantom{\sum}}^{\!\prime}
\sum_{n=1}^{\infty}\frac{1}{\sqrt{n}}
\nonumber\\
&&~
\times
\int_{\zeta_l}^{\infty}\!\!\!\!v^{3/2}dv
(r_{\rm TM}^{2n}+r_{\rm TE}^{2n})
e^{-nv}{I}_1\left(\frac{A_z}{a}nv\right).
\label{eqN7}
\end{eqnarray}
\noindent
This is the general expession for the shift of the second power of
the oscillator frequency due to the Casimir force between an
elliptic cylindrical lens and a plate obtained using the PFA.
Equation (\ref{eqN7}) takes into account nonlinearity of the
oscillator. By making the replacement (\ref{eq39}) it can be
generalized to the case of an asymmetric cylindrical lens
consisting of the parts of two dissimilar elliptic cylinders.
Multiplying the right-hand side of Eq.~(\ref{eqN7}) by the
factor $G(A,B;\varphi)$ defined in Eq.~(\ref{eq34}) one
obtains the generalization of Eq.~(\ref{eqN7}) for the case
of elliptic cylindrical lens rotated through an angle
$\varphi$ (see Sec.~IV).

It is easily seen that in the linear approximation Eq.~(\ref{eqN7})
leads to familiar expressions commonly used in the literature
\cite{2,3,15,16,35,36,37,38}. In fact the linear regime of the
oscillator, and the first nonlinear corrections to it considered
in Ref.~\cite{36}, are obtained by using the following
representation for the Bessel function \cite{45}
\begin{equation}
I_1(z)=\sum_{k=0}^{\infty}
\frac{\left(\frac{z}{2}\right)^{2k+1}}{k!(k+1)!}=
\frac{z}{2}+\frac{z^3}{16}+ O(z^5).
\label{eqN8}
\end{equation}
\noindent
Substituting the first term on the right-hand side of this
equation into  Eq.~(\ref{eqN7}) and performing the summation over
$n$, one finds
\begin{eqnarray}
&&
\omega_r^2-\omega_0^2=-C\frac{k_BTL}{4\sqrt{\pi}a^3}\,\frac{A}{\sqrt{2aB}}
\sum_{l=0}^{\infty}{\vphantom{\sum}}^{\!\prime}
\int_{\zeta_l}^{\infty}\!\!\!\!v^{5/2}dv
\nonumber\\
&&~
\times
\left[{\rm Li}_{-1/2}(r_{\rm TM}^{2}e^{-v})+
{\rm Li}_{-1/2}(r_{\rm TE}^{2}e^{-v})\right].
\label{eqN10}
\end{eqnarray}
\noindent
We can now use the Lifshitz-type formula (\ref{eq16}) for the gradient
of the Casimir force, and arrive at \cite{2,3,15,16,35,36,37,38}
\begin{equation}
\omega_r^2-\omega_0^2=-C\frac{\partial F(a,T)}{\partial a}.
\label{eqN11}
\end{equation}
\noindent
Bearing in mind that $\omega_r\approx\omega_0$ and, thus,
$\omega_r+\omega_0\approx 2\omega_0$, Eq.~(\ref{eqN11}) is
often presented in the form
\begin{equation}
\omega_r=\omega_0\left[1-\frac{C}{2\omega_0^2}\,
\frac{\partial F(a,T)}{\partial a}\right].
\label{eqN12}
\end{equation}
\noindent
In the linear regime, Eqs.~(\ref{eqN11}) and (\ref{eqN12})
allow calculation of the frequency shift of the oscillator
due to the Casimir force between a plate and an elliptic
cylinder. However, beyond the linear regime, the frequency
shift should be calculated using Eq.~(\ref{eqN7}).
This allows reliable comparison between the experimental
data and theoretical results for the Casimir force.

\section{Conclusions and discussion}

In this paper we have investigated the Casimir force acting
between a plate and a microfabricated elliptic cylindrical lens
made of real materials. This problem is of topical interest
from the experimental point of view. The
application of available technologies discussed by us in Sec.~II
leads to a fabrication of elliptic cylindrical lenses on
top of a micromachined oscillator, rather than
just the circular
cylindrical lenses considered previously in the literature.
We have obtained the Lifshitz-type formulas for the Casimir
force and for the gradient of the Casimir force in the
configuration of an elliptic cylinder and a plate.
In the framework of the PFA (which is applicable at separations
much less than the smaller semiaxis of an elliptic cylinder),
the results for an elliptic cylinder are obtained from the
respective results for a circular cylinder by replacing the
cylinder radius $R$ with $A^2/B$, where $A$ and $B$ are the
semiaxes of an elliptic cylinder (see Sec.~III).

Bearing in mind that nanotechnological fabrication procedures
may lead to cylinders with deviations from perfect elliptic
shape, we considered two types of such deviations.
In Sec.~IV we obtained the Lifshitz-type formulas for the
Casimir force and for its gradient in the configuration of
a plate near asymmetric elliptic cylindrical lenses.
Specifically, the constraints on an admissible angle of
rotation of an elliptic cylindrical lens about the
cylinder axis were found, allowing sufficiently small
deviations from the values of the Casimir force and its
gradient computed for the case of zero rotation angle.
The respective results for both perfect and asymmetric
elliptic cylindrical lenses were also obtained for the
electrostatic force in plate-lens configuration used for
calibration purposes in experiments on measuring the
Casimir force.
Note also that corrections to the Casimir force and its
gradient due to nonparallelity of a plate and an elliptic
cylinder are approximately the same as in the case of
circular cylinder considered in Ref.~\cite{33}.

For the needs of several proposed experiments on measuring the
Casimir force in a cylinder-plate geometry, we have considered
an oscillator with an attached elliptic cylindrical lens
interacting with the plane plate both made of real materials.
For dynamic measurements, when the separation distance between
a lens and a plate is varied harmonically, we have found the
frequency shift of an oscillator due to the Casimir force
in the nonlinear regime (Sec.~V). The resulting equations can be
used at short separations between a lens and a plate
where the commonly used linear equations are not applicable.
At the same time, it is shown that in the linear approximation
our result yields to the known expression.

To conclude, the proposed experiment on measuring the Casimir
force between a microfabricated elliptic cylindrical lens on
the top of a micromachined oscillator and a plate is of much
current interest and can shed additional light on the problem
of thermal Casimir force.

\section*{Acknowledgments}
R.S.D.~acknowledges NSF support through Grant No.\ PHY--0701236
and LANL support through contract No.\ 49423--001--07.
D.L.\ and R.S.D.\ acknowledge support from DARPA grant No.\ 09--Y557.
E.F. was supported in part by DOE under Grant No.~DE-76ER071428.
G.L.K.\ and V.M.M.\ are grateful to the Department of Physics,
Purdue University for financial support.
G.L.K.\ was also partially supported by the Grant of the Russian
Ministry of Education P--184.

\newpage
\begin{figure*}[h]
\vspace*{-6.cm}
\centerline{\hspace*{2cm}
\includegraphics{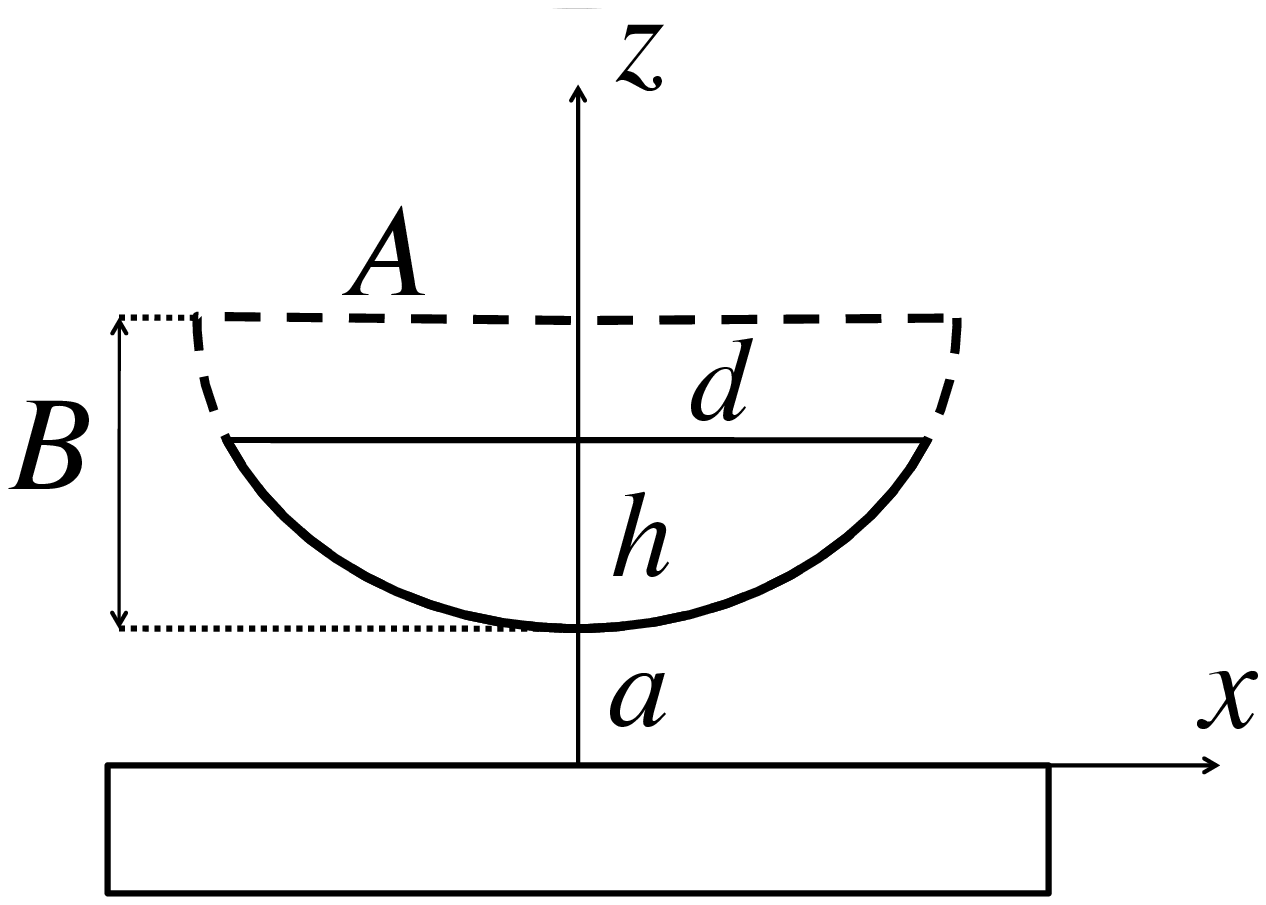}
}
\vspace*{-9.5cm}
\caption{Elliptic cylindrical lens of
thickness $h$ and width $2d$
obtained from an elliptic cylinder with
semiaxes $A$ and $B$
above a plate. The figure is not to scale.
}
\end{figure*}
\begin{figure*}[h]
\vspace*{-4.cm}
\centerline{\hspace*{2cm}
\includegraphics{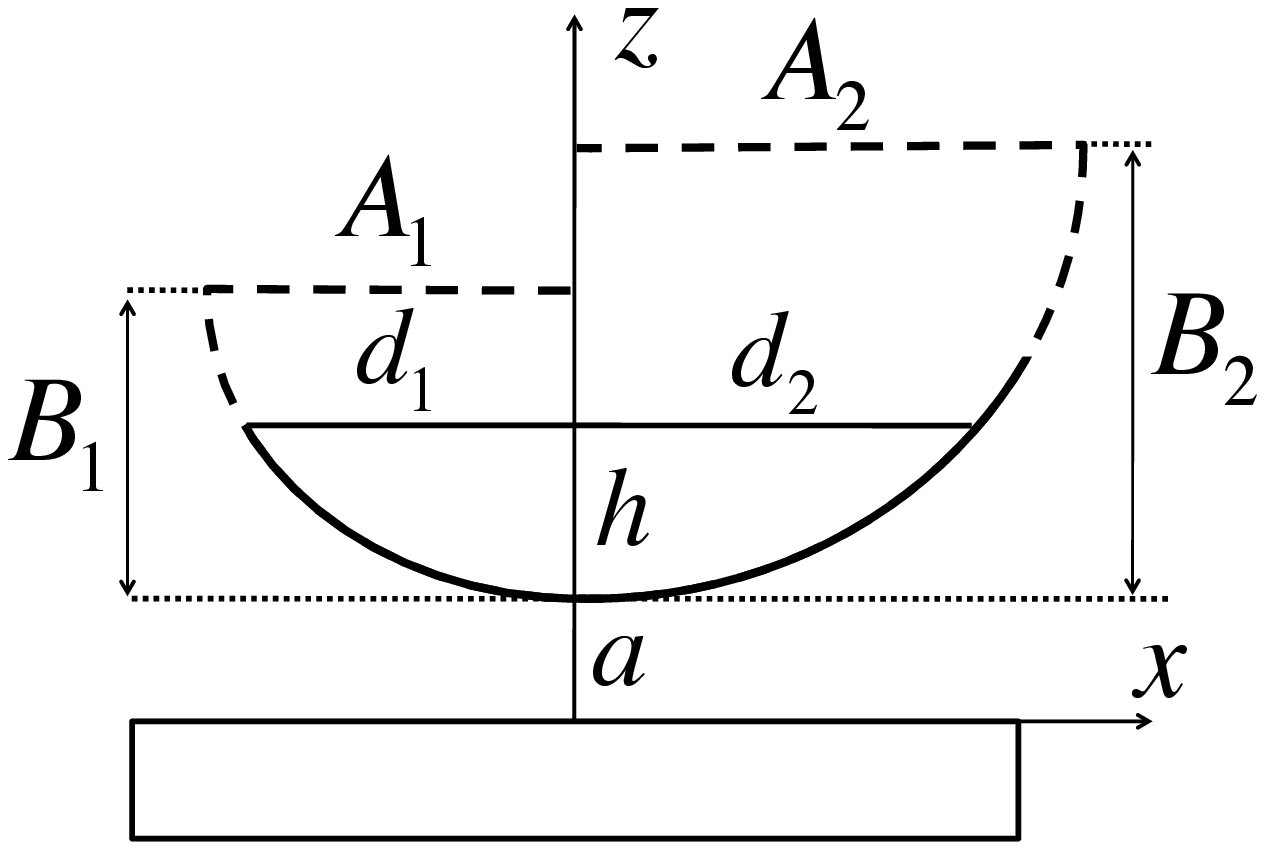}
}
\vspace*{-9.5cm}
\caption{An asymmetric elliptic cylindrical lens of
thickness $h$ and width $d_1+d_2$
obtained from two elliptic cylinders with
semiaxes $A_1,\,B_1$ and $A_2,\,B_2$
above a plate. The figure is not to scale.
}
\end{figure*}
\begin{figure*}[h]
\vspace*{0.cm}
\centerline{\hspace*{2cm}
\includegraphics{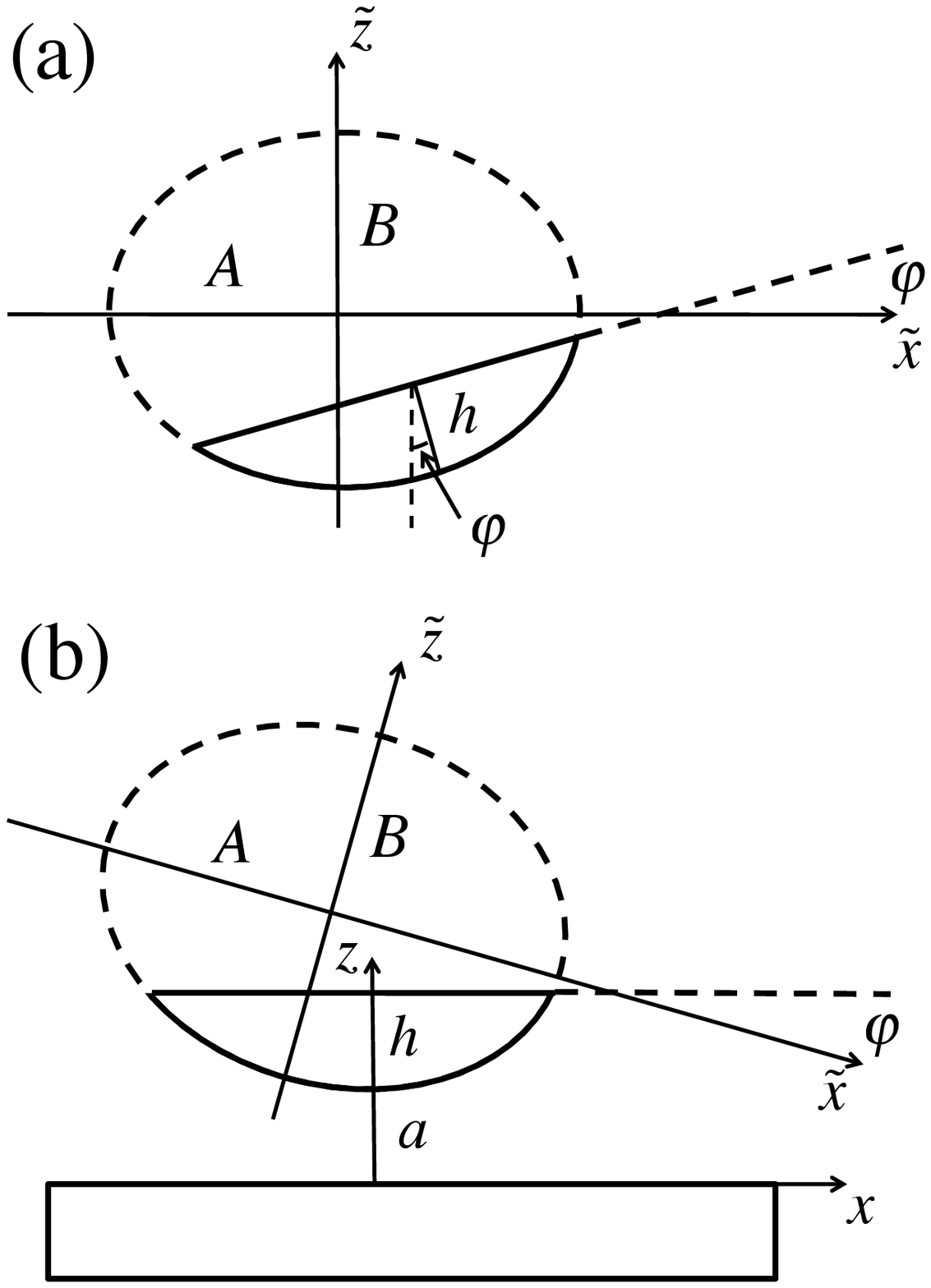}
}
\vspace*{-11.5cm}
\caption{(a) An asymmetric elliptic cylindrical lens of
thickness $h$ obtained from an elliptic cylinder with
semiaxes $A$ and $B$. (b) The same asymmetric elliptic
cylindrical lens spaced at the closest separation $a$
above a plate. The figure is not to scale.
}
\end{figure*}
\begin{figure*}[h]
\vspace*{-6.cm}
\centerline{\hspace*{2cm}
\includegraphics{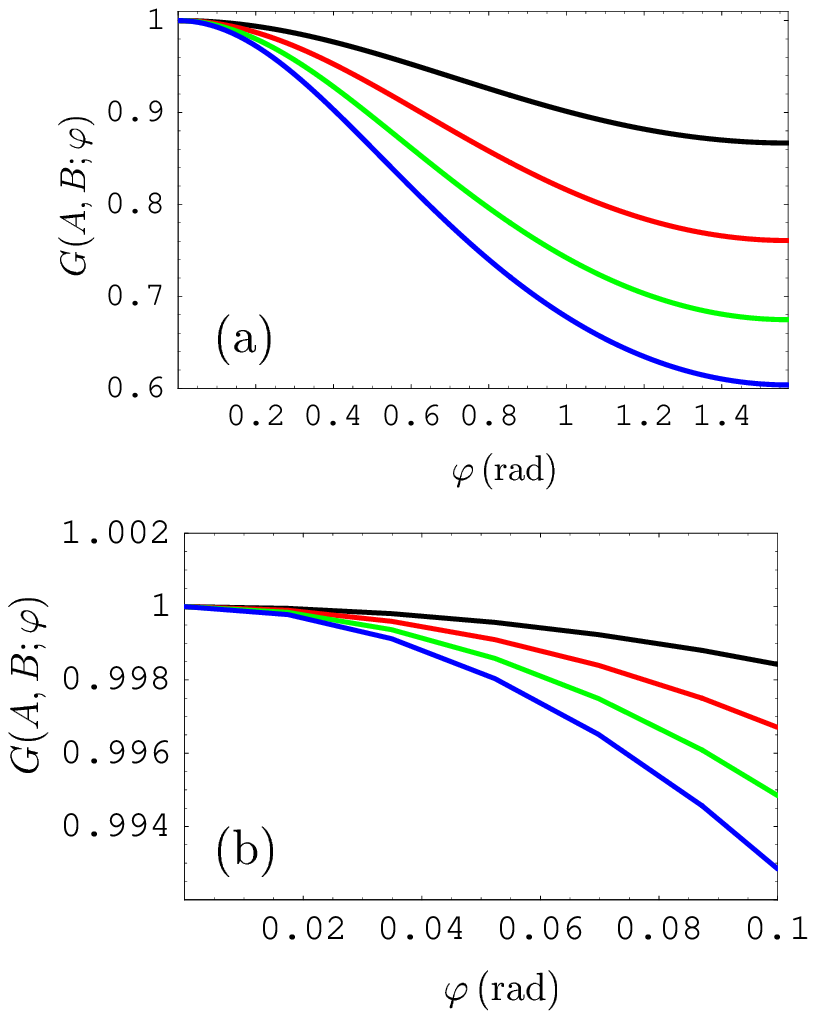}
}
\vspace*{-11.cm}
\caption{(Color online)
The ratio of the Casimir forces between an asymmetric elliptic
cylindrical lens and a plate  for the angle
of rotation equal to $\varphi$ and to zero as a function
of $\varphi$. For different lines the ratio of cylinder
semiaxes $A/B=1.1$, 1.2, 1.3, and 1.4 increasing from the top
to bottom lines. The interval of the angles of rotation
varies (a) from $\varphi=0$ to $\varphi=\pi/2$ and (b)
from $\varphi=0$ to $\varphi=0.1\,$rad.
}
\end{figure*}
\end{document}